\documentclass[a4paper,11pt]{article}

\usepackage{jheppub} 

\usepackage[T1]{fontenc} 

\usepackage{graphicx}
\usepackage{epsfig}
\usepackage{rotating}
\usepackage{amssymb}
\usepackage{subfigure}
\usepackage{dsfont}
\usepackage{psfrag}
\usepackage{amsmath,euscript,array,mathrsfs}
\usepackage{axodraw}
\usepackage{bbold}
\usepackage{epsf}

\newcommand{\beq}{\begin{equation}}
\newcommand{\eeq}{\end{equation}}
\newcommand{\beqs}{\begin{eqnarray}}
\newcommand{\eeqs}{\end{eqnarray}}

\newcommand{\Tr}{{\rm Tr}}

\def\hbar{\hspace{0pt}\raisebox{1pt}{$-$} \hspace{-7pt} h}

\newcommand{\be}{\begin{equation}}
\newcommand{\ee}{\end{equation}}

\newcommand{\bea}{\begin{eqnarray}}
\newcommand{\eea}{\end{eqnarray}}

\def\lbldef#1#2{\expandafter\gdef\csname #1\endcsname {#2}}

\def\href#1#2{#2}

\newcommand{\ber}{\begin{eqnarray}}
\newcommand{\eer}{\end{eqnarray}}

\newcommand{\beqar}{\begin{eqnarray}}

\newcommand{\eeqar}{\end{eqnarray}}

\newcommand{\dsl}
  {\kern.06em\hbox{\raise.15ex\hbox{$/$}\kern-.56em\hbox{$\partial$}}}

\newcommand{\eeqarr}{\end{eqnarray}}
\newcommand{\ZZ}{{\rm \kern 0.275em Z \kern -0.92em Z}\;}

\def\CC{{\mathchoice
{\rm C\mkern-8mu\vrule height1.45ex depth-.05ex
width.05em\mkern9mu\kern-.05em}
{\rm C\mkern-8mu\vrule height1.45ex depth-.05ex
width.05em\mkern9mu\kern-.05em}
{\rm C\mkern-8mu\vrule height1ex depth-.07ex
width.035em\mkern9mu\kern-.035em}
{\rm C\mkern-8mu\vrule height.65ex depth-.1ex
width.025em\mkern8mu\kern-.025em}}}

\def\RR{{\rm I\kern-1.6pt {\rm R}}}

\def\ZZ{{\rm Z}\kern-3.8pt {\rm Z} \kern2pt}
\def\IB{\relax{\rm I\kern-.18em B}}
\def\ID{\relax{\rm I\kern-.18em D}}
\def\II{\relax{\rm I\kern-.18em I}}
\def\IP{\relax{\rm I\kern-.18em P}}

\newcommand{\bear}{\begin{eqnarray}}
\newcommand{\eear}{\end{eqnarray}}

\def\6{\partial}

\def\bea{\begin{eqnarray}}
\def\eea{\end{eqnarray}}

\def\beqx{\begin{displaymath}}
\def\eeqx{\end{displaymath}}

\newcommand{\bmat}{\left(\begin{array}}
\newcommand{\emat}{\end{array}\right)}

\def\bo{{\raise-.3ex\hbox{\large$\Box$}}}               
\def\face{{\raise.2ex\hbox{$\displaystyle \bigodot$}\mskip-2.2mu \llap {$\ddot
        \smile$}}}                                   
\def\>{\rangle}                                      
\def\<{\langle}                                      

\def\leftrightarrowfill{$\mathsurround=0pt \mathord\leftarrow \mkern-6mu
        \cleaders\hbox{$\mkern-2mu \mathord- \mkern-2mu$}\hfill
        \mkern-6mu \mathord\rightarrow$}        
\def\dvec#1{\vbox{\ialign{##\crcr
        \leftrightarrowfill\crcr\noalign{\kern-1pt\nointerlineskip}
        $\hfil\displaystyle{#1}\hfil$\crcr}}}           
\def\Tr{{\rm Tr \,}}                                    

\def\-{\hphantom{-}}

\title{Analysis of a Dilaton EFT for Lattice Data}

\author[a]{Thomas Appelquist}

\affiliation[a]{Department of Physics, Sloane Laboratory, Yale University,\\Prospect Street, New Haven,Connecticut 06520, USA}

\author[a]{James Ingoldby}

\author[b]{Maurizio Piai}

\affiliation[b]{Department of Physics, College of Science, Swansea University,\\Singleton Park, Swansea, Wales, UK}

\date{\today}

\abstract{
In a recent paper, we developed and applied a dilaton-based effective field theory (EFT) to the analysis of lattice-simulation data for a class of confining gauge theories with near-conformal infrared behavior. It was employed there at the classical level to the $SU(3)$ gauge theory with eight Dirac fermions in the fundamental representation. Here, we explore the structure of the EFT further. We examine its application to lattice data (newly updated) for the $SU(3)$ theory with eight Dirac fermions in the fundamental representation, and the $SU(3)$ theory with two Dirac fermions in the sextet representation. In each case, we determine additional fit parameters and discuss uncertainties associated with extrapolation to zero fermion mass. We highlight universal features, study the EFT at the quantum loop level and discuss the importance of future lattice simulations.}

\begin{document}
\maketitle
\flushbottom



\section{Introduction}
\label{Sec:Introduction}

In recent years, improved lattice techniques and the availability of computational resources have allowed the study of strongly coupled gauge theories
that differ qualitatively from QCD. These studies have indicated that infrared conformal behavior rather than confinement appears when the number of massless fermions $N_f$ exceeds a critical value $N_{fc}$ \cite{DeGrand:2015zxa}. This number defines the bottom of the conformal window. Also, as $N_{fc}$ is approached from below, a remarkably light scalar particle appears in the spectrum of several lattice simulations. These include simulations of $SU(3)$ gauge theories with $N_f = 8$ Dirac fermions in the fundamental representation~\cite{LSD,LatKMI,LatKMI2} and with $N_f = 2$ Dirac fermions in the $2$-index symmetric (sextet) representation ~\cite{FHKNSW,FHKMNW,FHKMNW2}.

The appearance of a light scalar has led to the suggestion that this particle should be interpreted as a dilaton, an approximate Nambu-Goldstone boson associated with the spontaneous breaking of dilatation symmetry. Lattice simulations are carried out for a range of finite fermion masses $m$. In the studies of Refs.~\cite{LSD,LatKMI,LatKMI2} and~\cite{FHKNSW,FHKMNW,FHKMNW2}, the range is such that the scalar mass is of the same order as the mass of a multiplet of pseudoscalars, approximate Nambu-Goldstone bosons (NGB's) associated with the spontaneous breaking of chiral symmetry. Both the scalar and the pseudoscalars are light compared to the other physical states. In the limit $m \rightarrow 0$, the pseudoscalar mass is expected to extrapolate to zero while the scalar mass should extrapolate to a small but finite value. We here use the term NGB's to refer only to the pseudoscalars.

The relative lightness of the scalar and NGB's in the lattice simulations suggests that they be treated via an effective field theory (EFT) with only these degrees of freedom. Several authors have begun this program \cite{EFTDilaton1,EFTDilaton2,EFTDilaton3,EFTDilaton4} building on early investigations \cite{C,LLB}. In Ref.~\cite{AIP}, we noted that the lattice data for $N_f = 8$ Dirac fermions in the fundamental representation~\cite{LSD,LatKMI,LatKMI2} can be fit employing such an EFT at only the classical level. In this paper, we extend our treatment of this EFT, exploring its features at both the classical and quantum levels and extending the comparison with lattice data to include $N_f = 2$ Dirac fermions in the $2$-index symmetric (sextet) representation ~\cite{FHKNSW,FHKMNW,FHKMNW2}. It is notable that a rather simple EFT employed at the classical level accurately captures the essential features of a large set of lattice data.

In Section~\ref{Sec:EFT}, we describe the ingredients of the EFT including the small explicit breaking of scale invariance through a weak dilaton potential, and discuss features of the EFT at the classical level. In Section~\ref{Sec:Lattice}, we compare the classical (tree-level) EFT to the lattice data, determining certain parameters of the theory and constraining the shape of the dilaton potential at large field strength. In Section~\ref{Sec:loops}, we discuss corrections to the tree-level EFT arising from the heavy states present in the lattice data and from quantum loop corrections computed within the EFT. In Section~\ref{Sec:conclusions}, we summarize our results, comment briefly on possible phenomenological applications and discuss open problems.

\section{Classical EFT}
\label{Sec:EFT}

\subsection{Ingredients}

To describe the light states appearing in lattice simulations, we employ an EFT consisting of the NGB's along with a description of a light singlet scalar consistent with its interpretation as a dilaton. The Lagrangian density takes the form 
\beqs 
{\cal L}&=&\frac{1}{2}\partial_{\mu} \chi \partial^{\mu} \chi \,+\,{\cal L}_{\pi} \, +\,{\cal L}_M  \,-\,V(\chi) \, , 
\label{Eq:L} 
\eeqs 
where $\chi$ is the real, scalar dilaton field. 

The term ${\cal L}_{\pi}$ is given by 
\beqs 
{\cal L}_{\pi}&=&\frac{f_{\pi}^2}{4}\left(\frac{\chi}{f_{d}}\right)^2 \,\Tr\left[\partial_\mu \Sigma (\partial^{\mu} \Sigma)^{\dagger}\right]\,,
\label{Eq:Lpi} 
\eeqs 
where the $\Sigma$ field describes the NGB's arising from the spontaneous breaking of chiral symmetry. It transforms as $\Sigma\rightarrow U_L \Sigma U_R^{\dagger}$, with $U_L$ and $U_R$ the matrices of $SU(N_f)_L$ and $SU(N_f)_R$ transformations, and satisfies the nonlinear constraint $\Sigma \Sigma^\dagger = \mathbb{I}$. It can be written as $\Sigma=\exp\left[2i \pi/f_{\pi} \right]$ where $\pi=\sum_a\pi^a T^a$. The dilaton field acts here as a conformal compensator. The parameter $f_{\pi}$ is the NGB decay constant describing the spontaneous breaking of chiral symmetry and $f_d$ is the vacuum value of the dilaton field describing the spontaneous breaking of dilatation symmetry. These are independent parameters, since in the underlying theory there can be condensates that break scale symmetry but not chiral symmetry. Still, we expect them to be similar in magnitude, set by the confinement scale of the underlying gauge theory.

For lattice-simulation purposes, chiral symmetry must be broken explicitly by the introduction of a fermion mass term of the form $m\bar{\psi}\psi$, with subsequent extrapolation to $m=0$. The effect of this mass can be captured by supplementing the EFT with the term
\begin{align}
\mathcal{L}_M= \frac{m^2_\pi f^2_\pi}{4}\left(\frac{\chi}{f_d}\right)^y \Tr \left[\Sigma + \Sigma^\dagger\right]\,,
\label{Eq:LM}
\end{align}
where $m_{\pi}^2 = 2 m B_{\pi}$, and $B_{\pi}$ is a constant.
The form of $\mathcal{L}_M$ is such that it breaks scale and chiral symmetries in the same way as the fermion-bilinear mass term in the underlying gauge theory~\cite{LLB}, with $y$ taken to be the scaling dimension of $\bar{\psi}\psi$. This is an RG-scale dependent quantity; in the present context it should be taken to be defined at scales above the confinement scale, where the gauge coupling varies slowly. It has been suggested that $y\approx 2$ at this scale \cite{CandG,Shrock}. We take $y$ to be a constant, but keep it as a free parameter to be fit to the lattice data.

Finally, a dilaton potential $V(\chi)$ describes the explicit breaking of conformal symmetry even in the limit $m_{\pi}^2 \rightarrow 0$. It has a minimum at some value $f_d > 0$, and we anticipate it to be shallow satisfying $m_d^2 \ll (4 \pi f_{d})^2$. Several proposed forms of the dilaton potential have appeared in the literature, for example \cite{EFTDilaton1,GGS}. However we do not adopt an explicit form, instead observing that some predictions of the EFT are form independent. We allow the lattice data to determine certain features of the potential.

\subsection{Scaling Features}

The term ${\cal L}_M$ generates a mass for the NGB's and contains a new scalar self-interaction. The full dilaton potential becomes 
\beqs
W(\chi) = V(\chi) - \frac{N_f m_{\pi}^2 f_{\pi}^2}{ 2} \left(\frac{\chi }{f_d}\right)^{y}\,.
\label{Eq:W}
\eeqs
This potential is minimized at some field strength $\chi = F_d\, (\geq f_d) $, depending on the magnitude of the chiral-symmetry breaking term, which is not restricted to being a small contribution to $W(\chi)$. 
$F_d$ is finite under the assumption that $V(\chi)$ increases more rapidly than $\chi^y$ at large $\chi$.

For any non-zero $m_{\pi}^2$, it is convenient to express the EFT in terms of $y$ and a set of quantities $F_d$, $M_d^2$, $F_{\pi}$, and $M_{\pi}^2$, which extrapolate to their corresponding lower-case parameters in the $m^2_\pi\rightarrow0$ limit. The mass $M_d^2$ is determined by the curvature of the full potential at its minimum.  The other two quantities, $F_{\pi}$ and $M_{\pi}^2$, are identified from ${\cal L}_{\pi}$\ and ${\cal L}_M$ by taking $\chi = F_d$ and properly normalizing the NGB kinetic term.
They are given in general by simple scaling formulae \cite{EFTDilaton1,AIP}:
\beqs
\label{Eq:scaling1}
\frac{F_{\pi}^2}{f_{\pi}^2}&=&\frac{F_{d}^2}{f_{d}^2}\,,\\ 
\frac{M_{\pi}^2}{{m}_{\pi}^2}&=&\left(\frac{F_{d}^2}{f_{d}^2}\right)^{y/2-1}\,.
\label{Eq:scaling2}
\eeqs
We assume that these expressions apply in the $m^2_\pi\rightarrow0$ limit as well as in the larger-$m_{\pi}^2$ case where the second term in $W(\chi)$ begins to dominate the destabilizing of the scale-symmetric vacuum. In this regime, which applies to much of the current lattice data, $F_{d}^2/f_{d}^2 \gg 1$, increasing with $m_{\pi}^2$.

In general, with the field redefinition $\chi \equiv F_d + \bar{\chi}$, we can express the EFT in terms of the capitalized quantities:
\beqs 
{\cal L}_{\pi}&=&\frac{F_{\pi}^2}{4}\left[ 1+ \frac{\bar{\chi}}{F_{d}}\right]^2 \,\Tr\left[\partial_\mu \Sigma (\partial^{\mu} \Sigma)^{\dagger}\right]\,,
\label{Eq:LpiR} 
\eeqs
and
\beqs
{\cal L}_M&=&\frac{M_{\pi}^{2}F_{\pi}^2}{4}\left[1+ \frac{\bar{\chi}}{F_{d}}\right]^y\, \left[\Tr \left(\Sigma + \Sigma^{\dagger}\right) - 2N_f \right] \,,\label{Eq:LMR}
\eeqs
where $\Sigma=\exp\left[2i \Pi/F_{\pi} \right]$ and $\Pi\equiv(F_\pi/f_\pi)\pi$. We have removed the piece from $\mathcal{L}_M$ that contributes to the full dilaton potential $W(\chi)$. This potential can be re-expressed in terms of $F_d$, $M_d$, $y$ and possible additional parameters entering $V$. As an expansion in $\bar{\chi} / F_d$, $W$  takes the form 
\beqs
W(\bar{\chi}) = \text{constant} + \frac{M_d^2}{2} \bar{\chi}^2 + \frac{\lambda}{3!} \frac{M_d^2}{F_d} \bar{\chi}^3 + \frac{\gamma \, M_d^2}{4!\,F_d^2}\bar{\chi}^4\,+\,\cdots \,,
\label{Eq:Wexpansion}
\eeqs
where $\lambda$ and $\gamma$ are dimensionless quantities depending on $y$, the large-$\chi$ form of $V$, and $m_{\pi}^2$. This form of the EFT, expressed in terms of the capitalized quantities which scale up with $m_{\pi}^2$, will be helpful in estimating the size of quantum loop corrections to the classical theory.

It is important to note that the EFT treats the pseudoscalar states described by the $\Pi$ field as (pseudo) NGB's even for the larger values of $m_{\pi}^2 = 2B_{\pi} m $, which apply to most of the current lattice data. This could eventually break down since at sufficiently large $m$, the explicit breaking of chiral symmetry in the underlying gauge theory becomes large. However, for the lattice data to be discussed in Section~\ref{Sec:Lattice}, $M_{\pi}^2$ is small relative to the scale $(4\pi F_\pi)^2$, indicating that the pseudoscalars maintain their NGB character.

\section{Comparison to Lattice Data} 
\label{Sec:Lattice}

\subsection{Preliminaries} 

We apply the EFT at the classical level to the lattice data from the LSD collaboration for the $SU(3)$ gauge theory with $N_f = 8$ fundamental Dirac fermions~\cite{LSD}\footnote{For this theory, we do not include the LatKMI data \cite{LatKMI,LatKMI2} in the present analysis. The lattice gauge coupling is different, and the lattice is smaller, leading to rather different systematic effects.}, and to data from LatHC collaboration for the $SU(3)$ gauge theory with $N_f = 2$ Dirac fermions in the $2$-index symmetric representation~\cite{FHKNSW,FHKMNW,FHKMNW2}. The collaborations have so far provided data for the quantities $F_{\pi}$, $M_{\pi}^2$, and $M_d^2$. Each is measured for a set of non-zero values of $m_{\pi}^2 = 2 m B_{\pi}$. By fitting the EFT to the lattice data, we test the EFT framework, determine properties of the dilaton potential $V(\chi)$, and compute values for $y$ and other parameters.

The quantity $F_\pi$, defined using the conventions of Ref.~\cite{LSD}, is obtained from lattice calculations of the two point correlation function of axial-vector currents. The quantity $F_d$ (the VEV of the $\chi$ field) has not yet been obtained from a lattice calculation of a gauge theory correlation function. The issue of how $F_d$ can be determined directly from such a correlation function requires more study. However, we do not need to have lattice data for $F_d$ in order to apply our analysis. We first use only the data for $F_\pi$ and $M^2_\pi$ as they have the smallest uncertainties, and afterwards we add the data for $M^2_d$.

While the determination of the parameters $f_{\pi}$, $f_d$, and $m_d$ requires extrapolation to the $m^2_\pi\rightarrow0$ limit, the parameter $y$ enters through the chiral symmetry breaking term $\mathcal{L}_M$, and can be extracted directly from the finite-$m_{\pi}^2$ data. The two scaling relations in Eqs.~(\ref{Eq:scaling1}) and (\ref{Eq:scaling2}) can be combined to give  
\beqs
M_{\pi}^2 F_{\pi}^{2-y} = C m \,,
\label{Eq:y}
\eeqs
where $C = 2 B_{\pi} f_{\pi}^{2-y}$, independently of the dilaton potential. Lattice data for $M_{\pi}^2$ and $F_{\pi}$ alone can determine $y$ accurately. 

The finite-$m_{\pi}^2$ data can also be used to constrain the large-$\chi$ behavior of the dilaton potential $V(\chi)$. From Eq.~(\ref{Eq:W}) and the scaling relations Eqs.~(\ref{Eq:scaling1}) and (\ref{Eq:scaling2}), we have 
\beqs
\left. \frac{\partial V}{ \partial \chi}\right|_{\chi = F_d} = \frac{yN_{f}m_{\pi}^2f_{\pi}^2}{2f{_d^y}} F_{d}^{y-1} = \frac{yN_{f}f_\pi}{2f{_d}} M_{\pi}^{2} F_{\pi}\, .
\label{Eq:dV}
\eeqs
This shows that lattice data for $M^2_\pi$ and $F_\pi$ can be used to determine the gradient of $V(\chi)$ at the field value $\chi=F_d$ up to a constant of proportionality. Noting that $F_d\propto F_\pi$, it can be seen that data for $M^2_\pi$ and $F_\pi$ alone can fix the functional form of the potential. We make use of this result to constrain the large-$\chi$ behavior of $V(\chi)$.

Lattice data for the dilaton mass $M_d^2$ can also be included. Doing so provides an independent determination of the second derivative of the potential $V(\chi)$. A simple exercise leads to ~\cite{AIP}:
\beqs
\left.\frac{\partial^2V}{\partial \chi^2}\right|_{\chi=F_d}&=&M_d^2+\frac{y(y-1)N_f f_{\pi}^2}{2f_d^2}M_{\pi}^2\,.
\label{Eq:ddV}
\eeqs
The errors on $M^2_d$ are currently large. Nevertheless, we will make use of this relation in Section \ref{Sec:potential} to obtain a determination of the ratio $f^2_\pi/f^2_d$. 

We take the $N_f=8$ data from Ref.~\cite{LSD}, and the sextet data from Refs.~\cite{FHKNSW,FHKMNW,FHKMNW2},
referring the reader to the original publications for technical details.
We use lattice data at finite lattice spacing $a$, without continuum extrapolation. 
The mass and decay constant of the NGB's, and the mass of the dilaton for the $N_f=8$ theory are reported in Fig.~\ref{Fig:dataLSD}. The same quantities are reported for the sextet theory in Fig.~\ref{Fig:dataKUTI}. We see that $M^2_\pi,\,M^2_d\ll1/a^2$ throughout the range of the data for both theories, indicating that lattice discretization effects are small.

\subsection{Analysis Using Only Data for the NGB's}
\label{Sec:data}

\begin{figure}[h]
\begin{center}
\includegraphics[height=4.0cm]{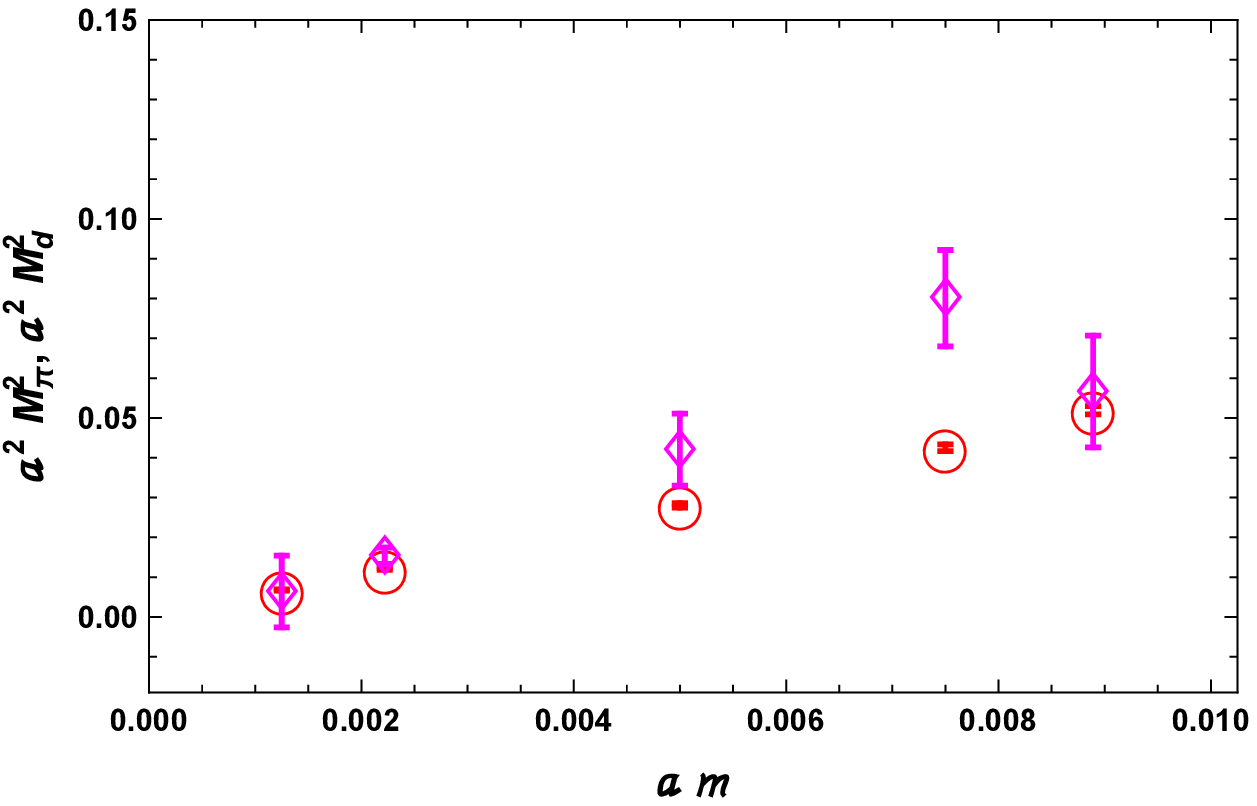}
\includegraphics[height=4.0cm]{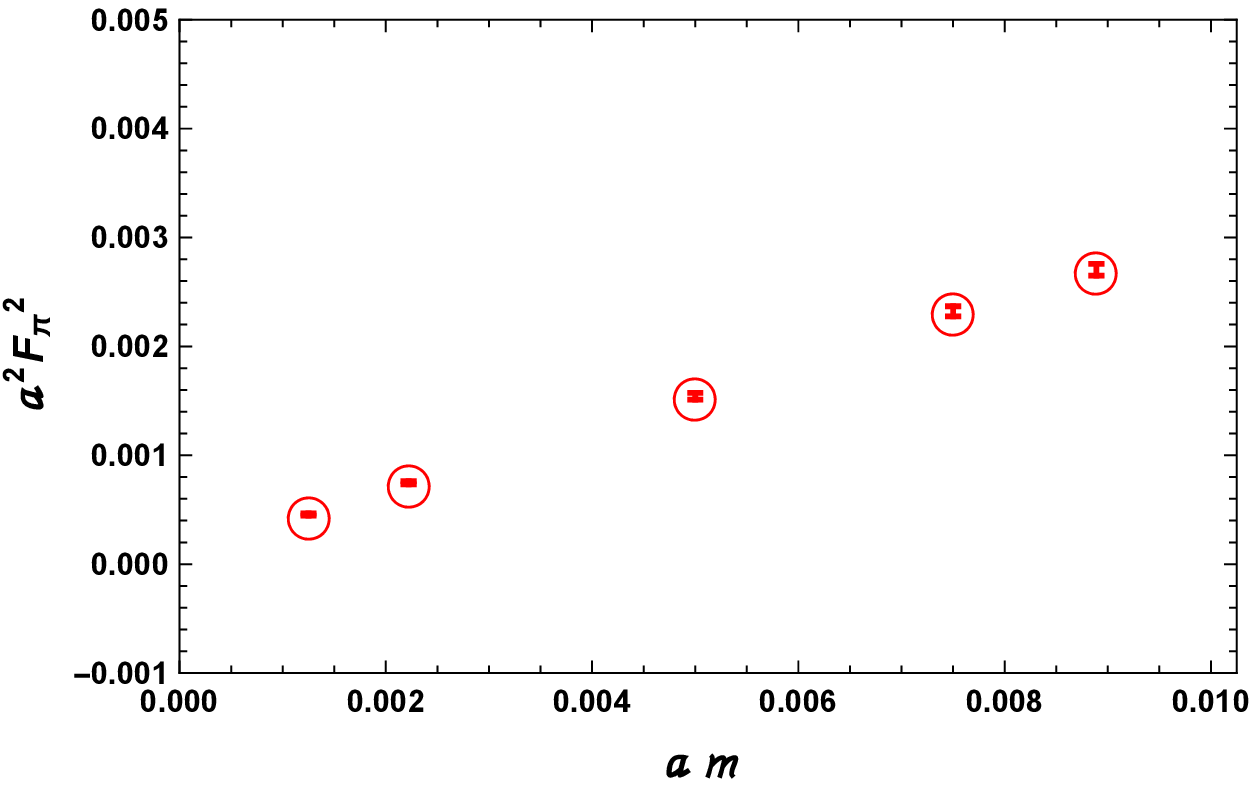}
\caption{Lattice data from the LSD collaboration for the $SU(3)$ theory with $N_f=8$ fundamentals~\cite{LSD}. Red circles represent the pseudoscalar data and their uncertainties are discussed in section~\ref{Sec:LSD}. Pink diamonds represent the scalar data with uncertainties discussed in section~\ref{Sec:potential}. The lattice spacing is denoted by $a$.}
\label{Fig:dataLSD}
\end{center}
\end{figure}

\begin{figure}[h]
\begin{center}
\includegraphics[height=4.0cm]{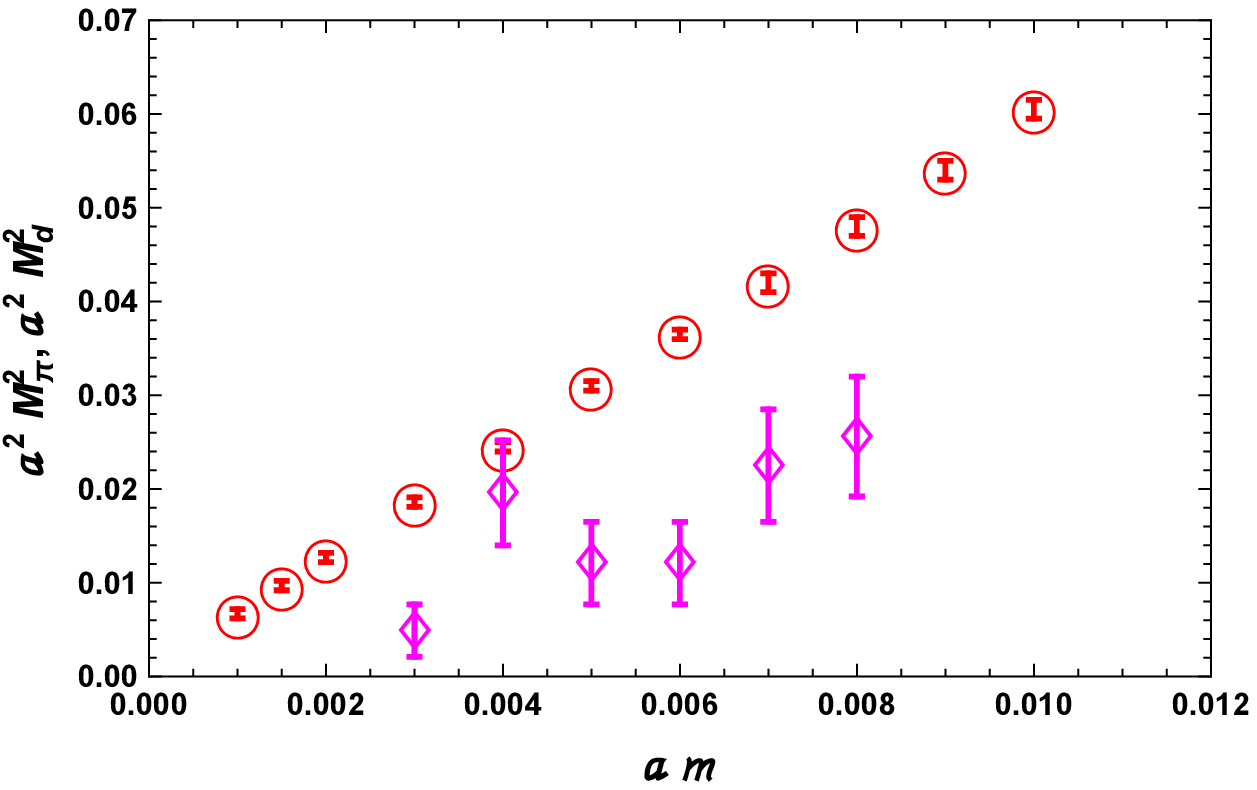}
\includegraphics[height=4.0cm]{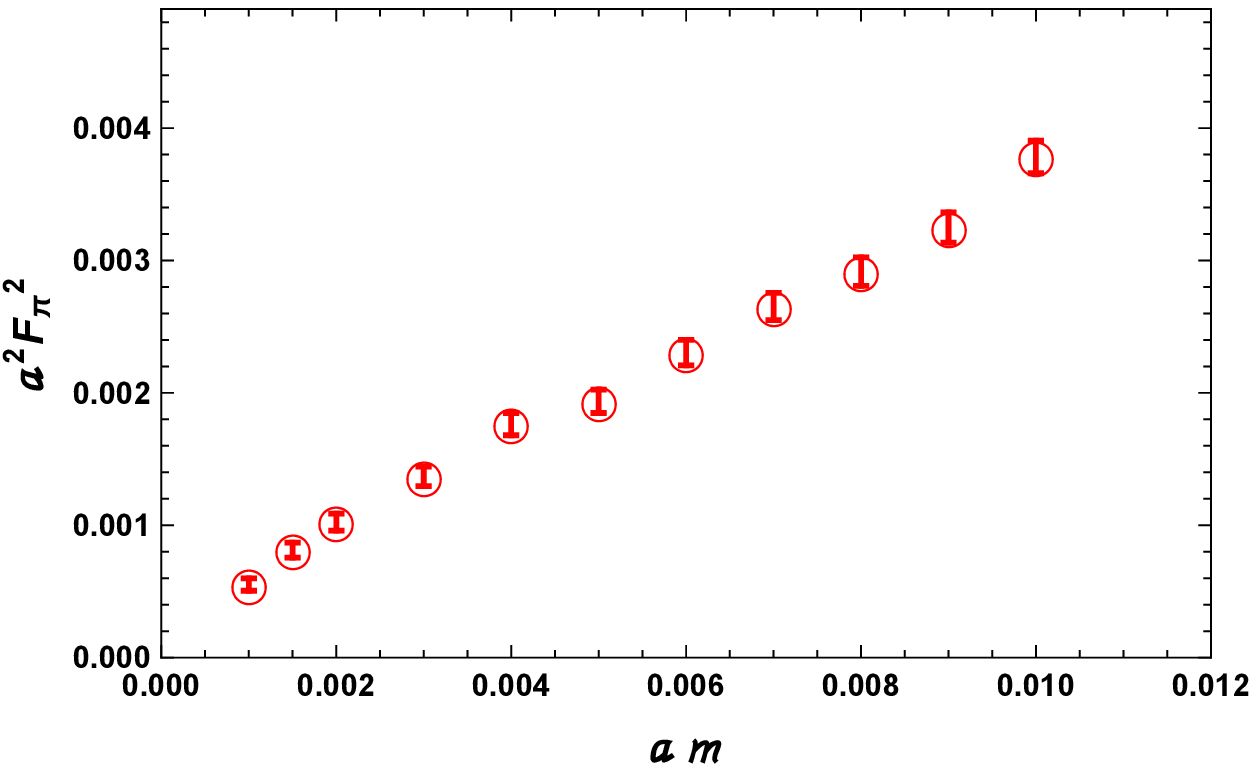}
\caption{Lattice data extracted from plots in Refs.~\cite{FHKNSW,FHKMNW,FHKMNW2} for the $SU(3)$ theory with $N_f=2$ sextets. Red circles represent the pseudoscalar data and pink diamonds represent the scalar. The lattice spacing is denoted by $a$. The errors are discussed in section~\ref{Sec:SU3}.}
\label{Fig:dataKUTI}
\end{center}
\end{figure}

We first utilize only the NGB data ($F_{\pi}^2$ and $M_{\pi}^2$), since it is currently more accurate than the $M_d^2$ data. We extract extracting from these two data sets values for $y$ and $C$ using Eq.~(\ref{Eq:y}) in a 2-parameter fit. We then make use of  Eq.~(\ref{Eq:dV}) to constrain the form of $V(\chi)$ for large $\chi$. This form, together with the scaling relation Eq.~(\ref{Eq:scaling1}), determines the relation between $F_{\pi}^2$ and $M_{\pi}^2$. The near linearity of both NGB data sets in Fig.~\ref{Fig:dataLSD} and Fig.~\ref{Fig:dataKUTI} implies that they are (approximately) linearly related to each other. Using Eq.~(\ref{Eq:dV}), it can therefore be seen that for large $\chi$, $V(\chi)$ behaves approximately like $\chi^4$.

We have kept open the form of $V(\chi)$, noting only that various proposals have appeared in the literature. The large-$\chi$ behavior in these proposals, which attempt to describe the scalar as a dilaton, typically include the power behavior $\chi^4$. This form is modulated by the factor log $\chi$ if the underlying conformal symmetry is broken by a marginal deformation \cite{GGS}. Here, we explore the constraint of the lattice data alone on the large-$\chi$ behavior of $V(\chi)$, by employing the simple phenomenological ansatz $V \propto \chi^p$. This form, while not in general theoretically based, is adequate to quantify the large-$\chi$ behavior of the potential, in particular its closeness to $\chi^4$. We obtain 
\beqs
M^2_\pi = B F_\pi^{^{\;\scriptstyle{p-2}}},
\label{Eq:pB}
\eeqs
where $B$ depends on the coefficient of $\chi^p$ in the potential. The potential will be well approximated by $\chi^p$ only at larger field strength, where the VEV satisfies $F_d \gg f_d$ and therefore $F_\pi\gg f_\pi$.

\subsubsection{$SU(3)$ with $N_f=8$}
\label{Sec:LSD}

\begin{figure}[h]
\begin{center}
	\subfigure{\includegraphics[height=4.8cm]{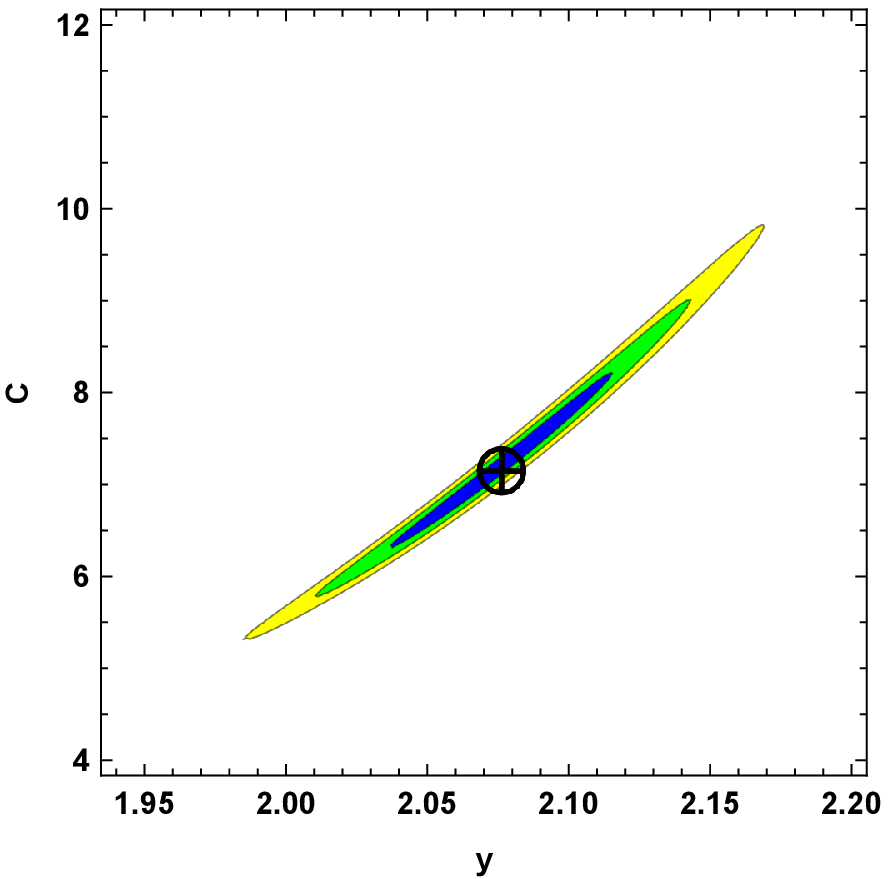}}\qquad
	\subfigure{\includegraphics[height=4.8cm]{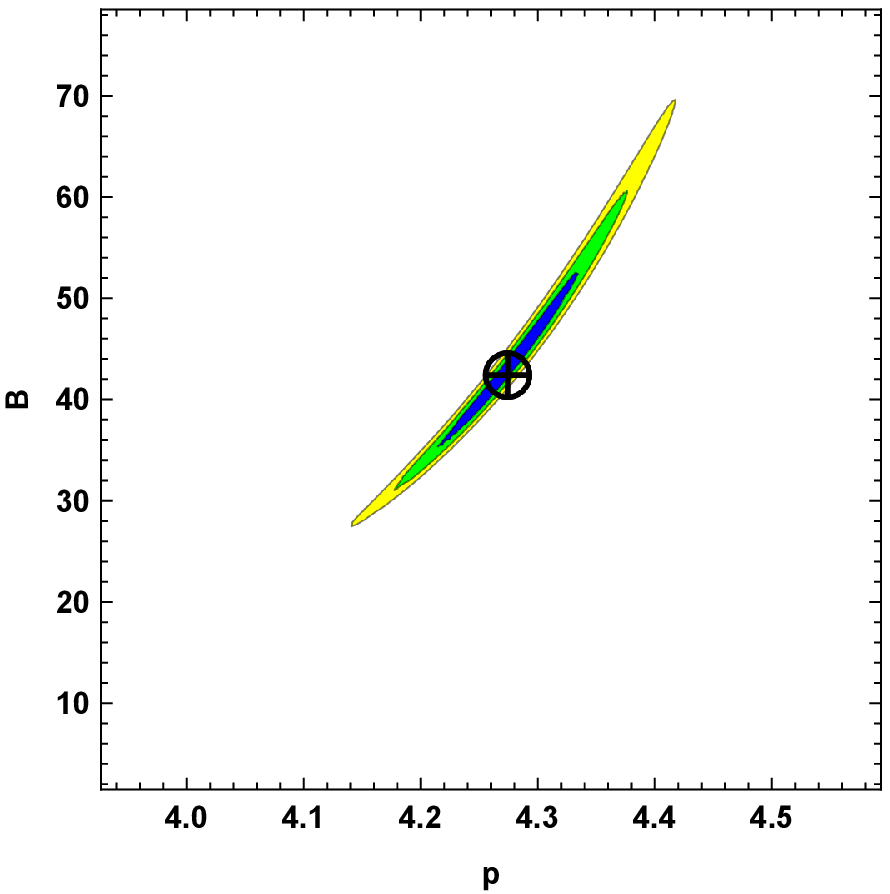}}
\caption{Contour plot from a 2-parameter fit based on Eq.~(\ref{Eq:y}) for the LSD data (left panel) and on Eq. (\ref{Eq:pB}), also for the LSD data (right panel). Contours correspond to $68.17\%$ c.l. (blue), $95.45\%$ c.l. (green) and $99.73\%$ c.l.,
obtained for $\Delta \chi^2=\{2.30\,,\,6.18\,,\,11.83\}$ respectively. The black crosses indicate the central values of the fit parameters.}
\label{Fig:yALSD}
\end{center}
\end{figure}

We first determine the parameters $y$ and $C$ from a fit of Eq.~(\ref{Eq:y}) to the LSD data. We use this fit equation in the form $(M_{\pi}a)^2 (F_{\pi}a)^{2-y} = C (ma)$ where $a$ is the lattice spacing, so that $C$ becomes a dimensionless number. The data can be obtained from the publicly available sources of Ref.~\cite{LSD}, the graphical displays there being reproduced in our Fig.~\ref{Fig:dataLSD}. For $F_{\pi}^2$ and $M_{\pi}^2$, the small error bars shown there are purely statistical. They can be seen to be smaller than $1\%$. Information about the correlation of these errors is not yet available publicly, and we do not take this into account. Since these quantities have been calculated without continuum extrapolation, there are larger, associated systematic errors. Drawing on the estimates in Ref.~\cite{LSD}, we therefore assign an overall, uncorrelated $2\%$ error to each of the $F_{\pi}^2$ and $M_{\pi}^2$ data points. The fit result is depicted in Fig.~\ref{Fig:yALSD}. The best-fit parameters are 
\beqs
y = 2.1\pm0.1\,,
\label{Eq:LSDy}
\eeqs
and $C = 7.2$, with $\chi^2/N_{dof}=0.34$. As noted in Ref.~\cite{AIP}, this result is consistent with $y=2$. Here and in the following, we take as a conservative indication of the global uncertainty the 99.73\% confidence level ranges obtained from our $\chi^2$ analysis by ignoring correlations (the extent of which are visible for example in Fig.~\ref{Fig:yALSD}). 
 
We next perform the 2-parameter fit based on Eq.~(\ref{Eq:pB}) to
determine the exponent $p$. We use this fit equation in the form $(M_{\pi}a)^2 = B(F_{\pi}a)^{p-2}$ so that $B$ becomes dimensionless. The best-fit parameters are 
\beqs
p = 4.3\pm0.2\, ,
\eeqs
and $B=43$ with $\chi^2/N_{dof} = 0.50$. The result of this fit is also shown in Fig. \ref{Fig:yALSD}. An additional systematic error can be assigned to this estimate stemming from the inclusion of the lowest mass points in the fit. These can be sensitive to the form of $V(\chi)$ for smaller $\chi$ where deviations from the simple form $\chi^p$ set in as the minimum of the potential is approached. There is some evidence in the lattice data for a deviation of this sort \cite{LSD}.  We have therefore carried out the fit also by eliminating one or two of the lowest mass points; in each case the central value of $p$ drops somewhat and the quoted statistical error grows somewhat. Because of this, we interpret our fit as being consistent with the behavior $V(\chi) \propto \chi^4$ at large $\chi$.

\subsubsection{$SU(3)$ with Sextets}
\label{Sec:SU3}

We next repeat the above exercise for the sextet theory. We draw on publicly available data, presented in graphical form in Refs.~\cite{FHKMNW,FHKMNW2,FHKNSW} and reproduced in our Fig.~\ref{Fig:dataKUTI}. For $a^2M_{\pi}^2$, we estimate the error on the lightest seven points to be approximately $0.0005$. The errors for the heavier four points, presented graphically in Ref.~\cite{FHKNSW} are larger. We conservatively take them to be $0.001$. Similarly, we estimate the error in $aF_{\pi}$ for each point to be $0.001$. We do not include systematic error estimates as they are not available in Refs.~\cite{FHKMNW,FHKMNW2,FHKNSW}. We note, though, that the errors we do include are of the same order as the systematic errors we included for the $N_f=8$ data. 

\begin{figure}[h]
\begin{center}
\subfigure{\includegraphics[height=4.8cm]{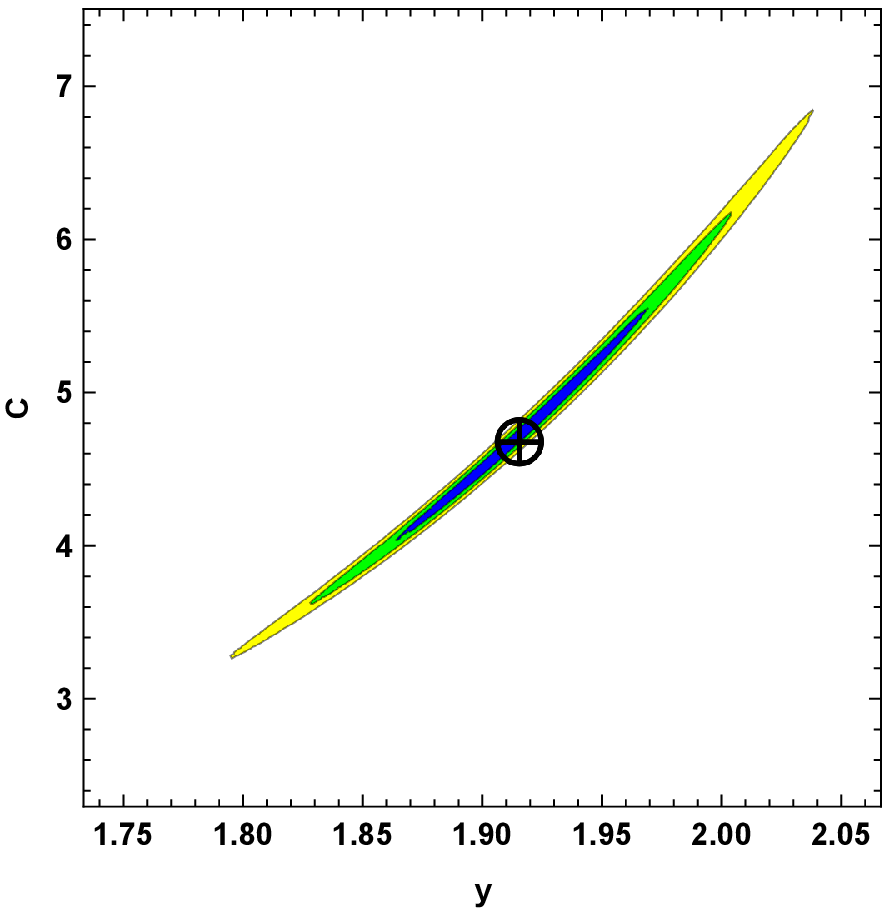}}\qquad
\subfigure{\includegraphics[height=4.8cm]{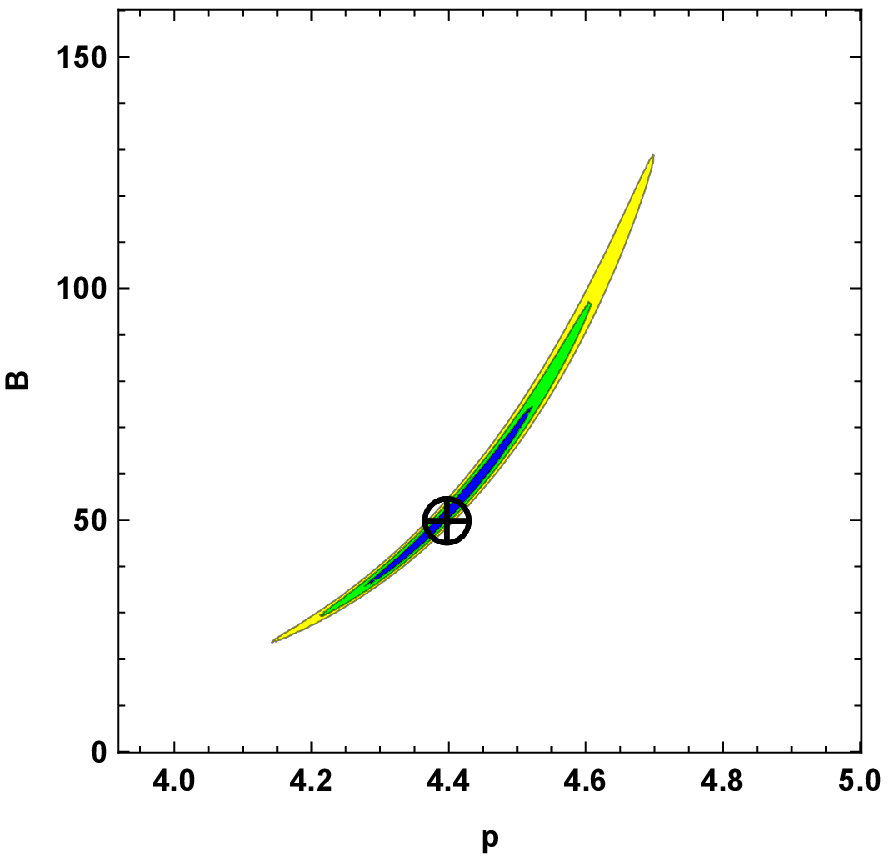}} 
\caption{Contour plot from the 2-parameter fit based on Eq.~(\ref{Eq:y}) for the sextet data (left panel) and on Eq.~(\ref{Eq:pB}) also for the sextet data (right panel). We show the contours corresponding to $68.17\%$ c.l. (blue), $95.45\%$ c.l. (green) and $99.73\%$ c.l.,
obtained for $\Delta \chi^2=\{2.30\,,\,6.18\,,\,11.83\}$ respectively. The black crosses indicate the central values of the fit parameters.}
\label{Fig:yAKUTI}
\end{center}
\end{figure}

We again use the fit equation $(M_{\pi}a)^2 (F_{\pi}a)^{2-y} = C (ma)$ and determine the best-fit parameters to be
\beqs
y = 1.9\pm0.1\,,
\label{Eq:Kutiy}
\eeqs
and $C=4.7$ with $\chi^2/N_{dof}=0.19$. The result is depicted in Fig.~\ref{Fig:yAKUTI}.

The result of the 2-parameter fit to the equation $(M_{\pi}a)^2 = B(F_{\pi}a)^{p-2}$ is also shown in Fig.~\ref{Fig:yAKUTI}. The best-fit parameters are 
\beqs
p=4.4\pm0.3\, ,
\label{Eq:pKuti}
\eeqs
 and $B=51$ with $\chi^2/N_{dof}=0.64$. Again, this fit should be regarded as being consistent with the behavior $V(\chi)\propto\chi^4$ at large $\chi$.\footnote{Following the initial posting of our paper, the authors of Refs.~\cite{FHKMNW,FHKMNW2,FHKNSW} posted a related paper \cite{Fodor:2017nlp}. We are pleased to note that the values they quote for the $y$ and $p$ parameters fall within the ranges of uncertainty given in Eqs.~(\ref{Eq:Kutiy}) and (\ref{Eq:pKuti}).}
\bigskip

Both lattice theories yield values of $y$ well below $3$ and compatible with $y=2$. This is not unexpected if $y$ is identified with the scaling dimension of $\bar{\psi}\psi$ in a confining gauge theory near the boundary of the conformal window. It is worth noting that the $N_f=8$ data lead to a central value of $y$ somewhat above $2$ while the converse is true for the sextet data. More precision would be needed to make a clearer statement about this specific point.

\subsection{Analysis Including the Scalar Mass Data}
\label{Sec:potential}

In this Section, we examine whether additional information can be gleaned from the existing lattice data, in particular about the extrapolated parameters $f_{\pi}^2$, $f_d^2$, and $m_d^2$. We already know from the scaling relation Eq.~(\ref{Eq:scaling1}) that within the framework of the EFT, the ratio $f_{\pi}^2/f_d^2$  can be directly determined if lattice data for $F_d^2$ become available to supplement the $F_{\pi}^2$ data. 
We show here that the ratio $f_{\pi}^2/f_d^2$ can be determined even in the absence of $F_d^2$ data by including the $M_d^2$ data in the fits. The errors are large for the $M_d^2$ data so this determination is currently limited in its accuracy.  Additional information about the parameters $f_{\pi}^2$, $f_d^2$, and $m_d^2$, for example the ratio $m_d^2/f_d^2$, will require data at smaller values of $m$.

\begin{figure}[h]
	\centering
	\includegraphics[height=4cm]{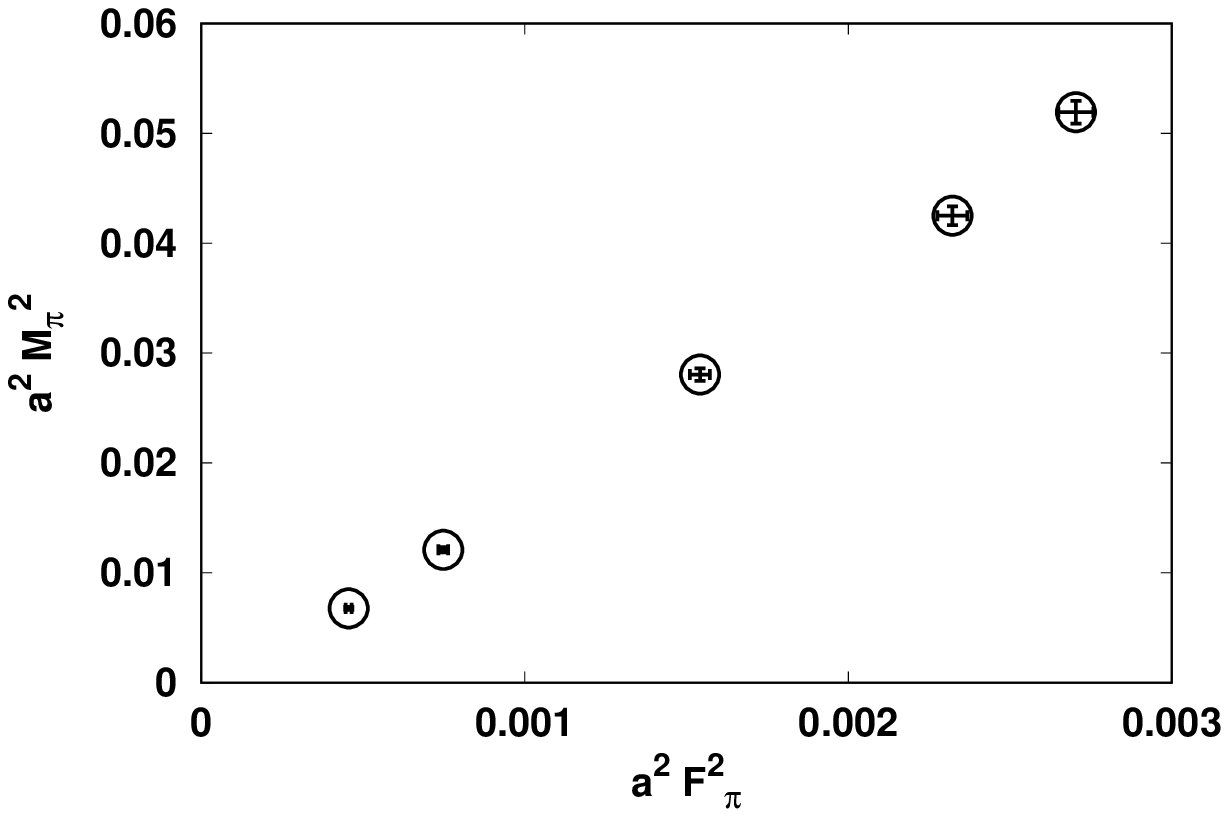} \qquad
	\includegraphics[height=4cm]{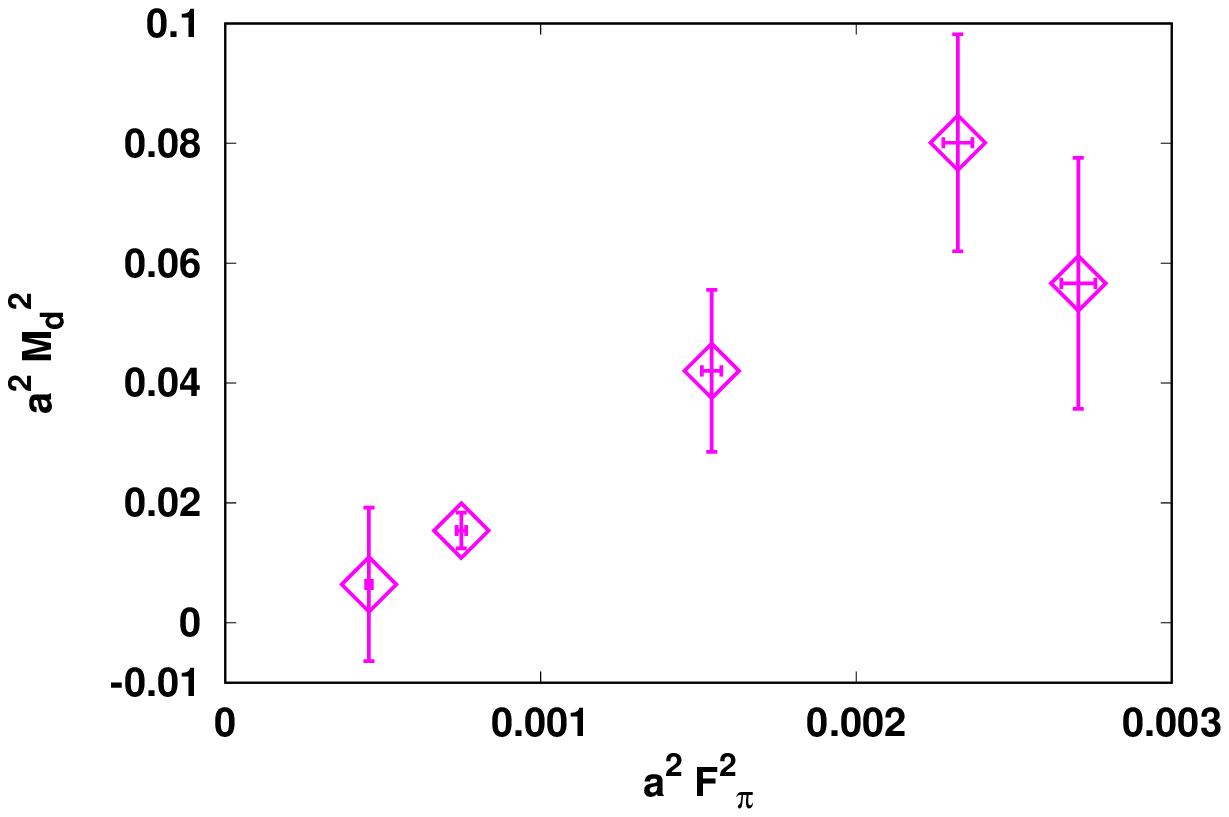}
	\caption{Lattice data for the $N_f=8$ theory. The error bars for $M^2_\pi$ and $F^2_\pi$ include an extra 2\% systematic error added to represent lattice artifacts \cite{LSD}. The $M^2_d$ errors are discussed in the text. The lattice spacing is denoted by $a$.}
	\label{Fig:M8}
\end{figure}

\begin{figure}[h]
	\centering
	\subfigure{\includegraphics[height=4cm]{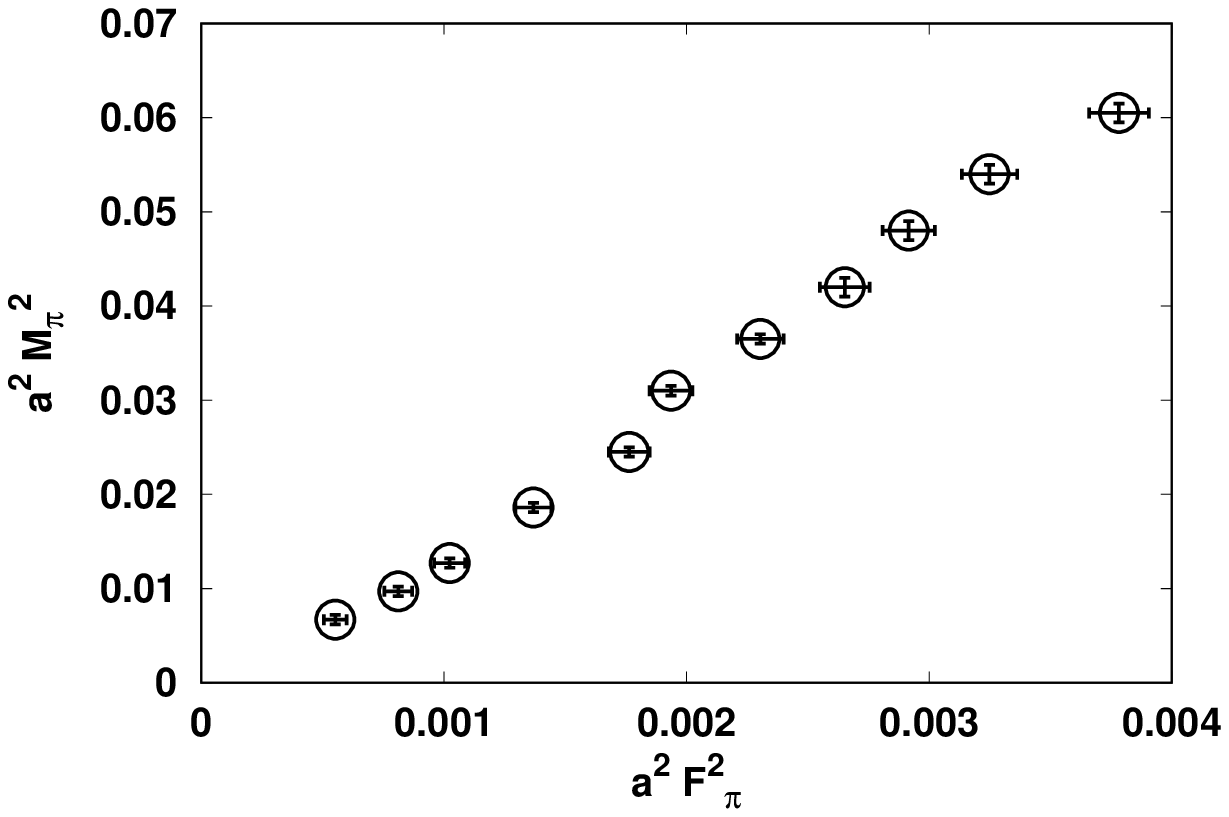}} \qquad
	\subfigure{\includegraphics[height=4cm]{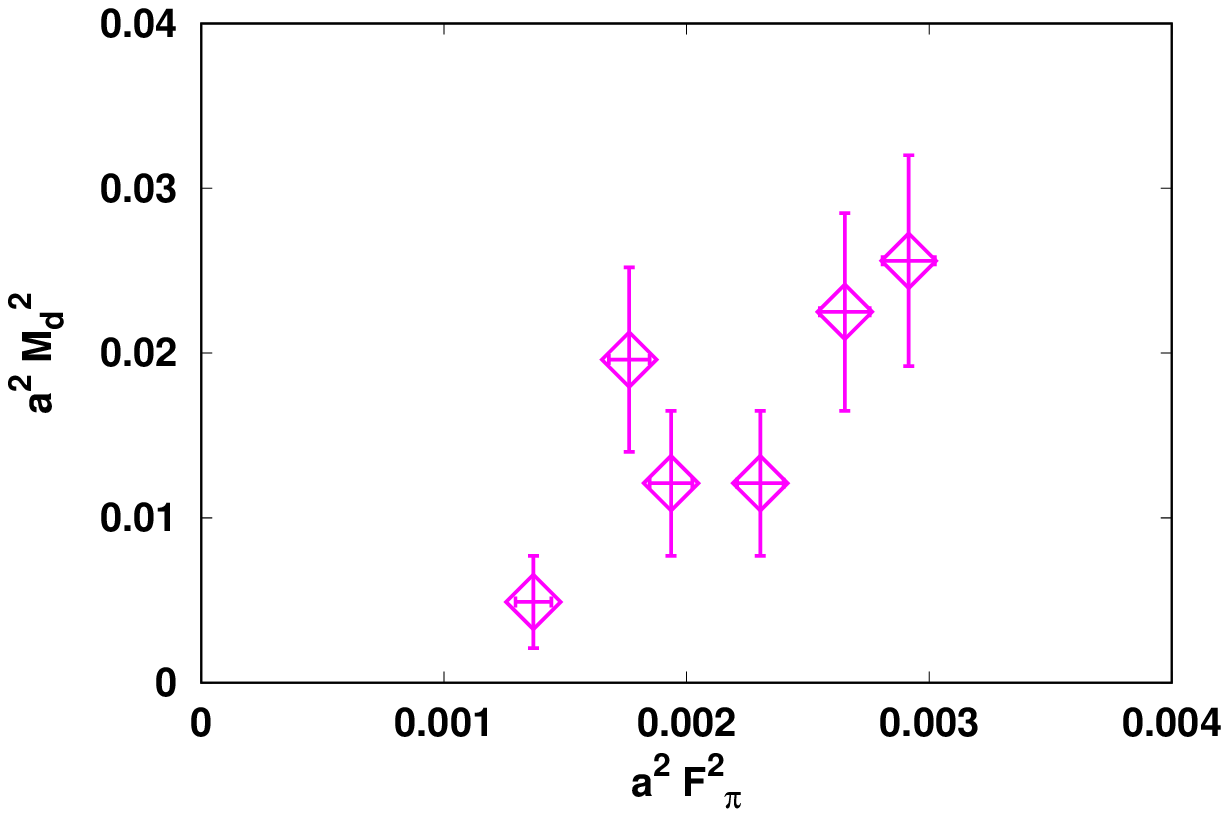}}
	\caption{Lattice data for the sextet theory, with errors, extracted from plots in Refs.~\cite{FHKMNW,FHKMNW2,FHKNSW}. The lattice spacing is denoted by $a$.}
	\label{Fig:MS}
\end{figure}

We use the scaling relation Eq.~(\ref{Eq:scaling1}) to recast Eqs.~(\ref{Eq:dV}) and (\ref{Eq:ddV}) into a form that shows the functional dependence of $M^2_\pi$ and $M^2_d$ on $F^2_\pi$ for any choice of the potential $V$.
\beqs
\label{Eq:MpiFpi}
M_{\pi}^2 & = & \frac{2f_d}{y N_{f} f_{\pi}} \frac{1}{F_\pi} V^{\prime}\left(\frac{f_d}{f_\pi}F_\pi\right)\, ,\\
M_{d}^2 & = & V^{\prime \prime}\left(\frac{f_{d}}{f_{\pi}}F_\pi\right)  - (y-1) \frac{f_\pi}{f_d}\frac{1}{F_\pi}     V^{\prime}\left(\frac{f_{d}}{f_{\pi}}F_\pi\right)\, .
\label{Eq:MdFpi}
\eeqs 
The prime and double-prime denote the first and second derivatives of $V$ with respect to its argument in parentheses. To fit Eqs.~(\ref{Eq:MpiFpi}) and (\ref{Eq:MdFpi}) to the lattice data shown in Figs.~\ref{Fig:M8} and \ref{Fig:MS}, we introduce an ansatz for the form of the potential as we did in Section~\ref{Sec:data}. Having already determined $y$ from the data, the two equations can then be employed together to determine $f^2_\pi/f^2_d$ as well as the shape of the potential for large $\chi$.

We noted already in Section~\ref{Sec:data} that the linearity of the data for $M_{\pi}^2$ versus $F_{\pi}^2$ implies through Eq.~(\ref{Eq:MpiFpi}) that $V(\chi)$ must behave approximately like $\chi^4$ for large $\chi$. We therefore repeat the type of fit employed there, using the phenomenological ansatz $V(\chi) \propto \chi^p$.  Eq.~(\ref{Eq:MpiFpi}) gives the fit equation (\ref{Eq:pB}) while Eq.~(\ref{Eq:MdFpi}) takes the form
\beqs
M^2_d = \frac{yN_ff^2_\pi}{2f^2_d}(p-y)BF_\pi^{^{\;\scriptstyle{p-2}}}\,.
\label{Eq:pBf}
\eeqs
Here, $B$ is the same quantity as in Eq. (\ref{Eq:pB}), and the new quantity $f_{\pi}^2/ f_d^2$ appears here. In our framework, this is a ratio of extrapolated quantities. Employing Eqs.~(\ref{Eq:pB}) and (\ref{Eq:pBf}) together, we have a three-parameter fit to the data for $M_{\pi}^2$ and $M_d^2$ versus $F_{\pi}^2$ with fit parameters $\{p,\; f_\pi^2/f_d^2,\; B\}$.


For the $N_f=8$ theory, performing the 3 parameter fit gives central values and errors for $p$ and $B$ consistent with those shown in Fig.~\ref{Fig:yALSD}. This is because the $M^2_d$ data have larger errors than the NGB data, and so provide a very weak additional constraint. The errors in $M_d$ are taken from Ref.~\cite{LSD} and include both statistical and some systematic effects. Using $y=2.1$, the fit then gives a central value and $99.73\%$ confidence interval for $f^2_\pi/f^2_d$:
\beqs
\frac{f^2_\pi}{f^2_d} = 0.08 \pm 0.04.
\label{Eq:Ratio8}
\eeqs
An estimate of the corresponding ratio in Ref.~\cite{LatKMI2} leads to a comparable result. The small uncertainty in $y$ contributes negligibly to the total uncertainty in $f^2_\pi/f^2_d$.

We similarly employ Eq.~(\ref{Eq:pBf}) to fit the sextet data, shown in Fig.~\ref{Fig:MS}. As noted earlier, we include errors for $M_{\pi}^2$ and $F_{\pi}^2$ estimated from the graphical display of data in Refs.~\cite{FHKMNW,FHKMNW2,FHKNSW}. The larger error bars for $M_{d}^2$ are taken from graphs in Ref.~\cite{FHKMNW}. The resultant values and errors of $p$ and $B$ are consistent with those shown in Fig.~\ref{Fig:yAKUTI}. Using $y=1.9$, the central value and $99.73\%$ confidence interval for $f^2_\pi/f^2_d$ given by the fit is
\beqs
\frac{f^2_\pi}{f^2_d} = 0.09 \pm 0.06.
\label{Eq:RatioS}
\eeqs

The ratio $f_{\pi}^2/f_d^2$ is well below unity for both theories, but the large uncertainties preclude a precise comparison of the two values. The determination of $f_{\pi}^2 / f_d^2$ does not depend strongly on the precise form of the ansatz $V\sim\chi^p$.  Any qualitatively similar ansatz that allows a behavior close to $\chi^4$ at large $\chi$, and therefore fits the data well, will yield values for $f_{\pi}^2 / f_d^2$ consistent with those that we quoted. While within our framework this ratio can be determined from lattice data at finite-$m$, this is not the case for the individual extrapolated quantities $m_d^2$, $f_d^2$ and $f_{\pi}^2$ or other ratios such as $m^2_d/f_d^2$.

Knowing $f^2_\pi/f^2_d$ allows us to predict the value of $F^2_d$ in the $N_f=8$ and sextet theories, using the scaling relation Eq.~(\ref{Eq:scaling1}). We find $F_d\sim3\,F_\pi$ in both theories for the range of fermion masses from the region of current lattice data to zero. We will use this result to estimate quantum loop corrections in Section~\ref{Sec:loops}.

\section {Heavy States and Quantum Loops }
\label{Sec:loops}

In Eqs.~(\ref{Eq:LpiR}) and (\ref{Eq:LMR}), 
we showed that $\cal{L}$, when expressed in terms of the capitalized (finite-$m_{\pi}^2$) quantities  $F_d$, $M_d^2$, $F_{\pi}$, and $M_{\pi}^2$, has the generic form 
\beqs
{\cal L}&=&\frac{F_{\pi}^2}{4}\left[ 1+ \frac{\bar{\chi}}{F_{d}}\right]^2 \,\Tr\left[\partial_\mu \Sigma (\partial^{\mu} \Sigma)^{\dagger}\right] + \frac{1}{2} \partial_\mu \bar{\chi} \partial^{\mu} \bar{\chi} + O(M_{\pi}^2) + O(M_{d}^2)\,,
\label{Eq:LCap}
\eeqs
where the final two terms are given by Eqs.~(\ref{Eq:LMR}) and (\ref{Eq:Wexpansion}). For any fixed value of $m_{\pi}^2 = 2 B_{\pi} m$, the cutoff on the EFT is of order $4 \pi F_{\pi}$. The EFT is expected to be weakly coupled if $M_{\pi}^2, M^2_d \ll (4 \pi F_{\pi})^2$ and $M_{\pi}^2, M^2_d \ll (4 \pi F_d)^2$ (and the energies are no larger than of order $M_{\pi} $ and $M_d$). In both the $N_f=8$ and sextet cases, we have found that $F_d>F_\pi$. From  inspection of Figs.~\ref{Fig:dataLSD} and~\ref{Fig:dataKUTI}, one can therefore see that these conditions on the masses are well satisfied throughout the range of $m$ values. The cutoff on the EFT scales up with $m$ in much the same way as the scalar and NGB masses, allowing for a finite range of validity for the EFT at all existing $m$ values.

In addition to the NGB's and dilaton scalar, lattice simulations of the $N_f=8$ and sextet theory show that other heavier states appear in the spectrum \cite{LSD,FHKMNW2}. Corrections to the use of the EFT at the classical level can be estimated by examining the effect of these states as well as the quantum loop diagrams that arise from Eq.~(\ref{Eq:LCap}). Both the heavy-state effects and the cutoff-dependent effects arising from the loops can be incorporated into the EFT through the addition of new operators, some of which are displayed in Ref.~\cite{EFTDilaton1}. Rather than enumerating these operators, we turn directly to estimating the contributions of heavy states and quantum loops to the lattice observables discussed in this paper.

\subsection{Heavy State Corrections}

Lattice data~\cite{LSD,FHKMNW2} show that for heavy states, the ratio $M_H^2/(4 \pi F_{\pi}) ^2$ is roughly constant as a function of $m$ throughout the existing range. Here $M_H$ represents any of the heavy masses. Furthermore, $M_d^2 \sim M_{\pi}^2 \ll M_H^2 \sim (4 \pi F_{\pi})^2$ throughout the range. The contributions to observables arising from heavy states are suppressed by powers of $M_{\pi}^2 / M_H^2$, $M_{d}^2 / M_H^2$, and $E^2 / M_H^2$, where $E$ is a typical energy of order $M_{\pi}$ or $M_{d}$. For the current LSD $N_f = 8$ data~\cite{LSD}, corrections of order $M_{\pi}^2 / M_H^2$ are $ \leq 0.3$, decreasing to $\leq 0.2$ for the lowest $m$ values. The data for $M_{d}^2 / M_H^2$, with their larger statistical errors, satisfy a similar bound. For the LatHC sextet data \cite{FHKMNW2}, the ratios $M_{\pi}^2 / M_H^2$ are also in this range, and $M_{d}^2 / M_H^2$ are smaller, again with larger statistical errors. For both theories, we expect that the ratios $M_{d}^2 / M_H^2$ will decrease to even smaller values in the $m\rightarrow0$ limit and that $M_{\pi}^2 / M_H^2$ will vanish.

\subsection{Quantum Loops} 

Since the EFT is relatively weakly coupled in the range of the lattice data, we anticipate that the quantum loop corrections are relatively small. Their computation is complicated by the role of scale symmetry. The weak scalar potential encodes the small breaking of scale symmetry at the classical level in the EFT, but this symmetry is naturally broken more strongly at the quantum level. Loop corrections, cut off at momentum scales of order $4 \pi F_{\pi}$ can lead to corrections to the potential of this order, requiring fine tuning, as with the Higgs boson mass in the minimal standard model. We do not address this issue directly here. We accept the weakness of the potential as indicated by the lattice data, dispensing with power-law sensitivity to the UV cutoff through the device of dimensional continuation.

This leaves a set of pole terms proportional to $1/\epsilon$ where $d = 4 -\epsilon$. Each signals a logarithmic sensitivity to the UV cutoff and therefore the generation of new, higher-dimension operators with unknown coefficients.  The logarithms are sensitive also to momenta on the order of $M_d$ and $M_{\pi}$ (chiral logarithms in the case of $M_{\pi}$) with their coefficients determined solely by the parameters in the tree-level Lagrangian.

We estimate the size of the one-loop corrections to the three observables for which we have lattice data by focusing on these logarithmic terms. Expressions for their contributions have appeared in the literature \cite{EFTDilaton3,EFTDilaton4,Bijnens:2009qm}. They have the generic form
\begin{align}
\frac{M^2}{(4\pi F)^2} \ln \left(\frac{M^2}{(4\pi F)^2}\right), \nonumber
\end{align}
 times known $O(1)$ coefficients and $N_f$-dependent counting factors, where $M^2$ and $F^2$ represent either NGB or scalar quantities. Such terms can be prominent because the logarithms are large and/or because the $N_f$ factors, which count the number of NGB's, are large. Here, the ratios $(4 \pi F)^2/M^2$ are not extremely large so we take the logarithms to be of order unity. For each observable, we then identify the largest term taking into account the $N_f$ dependence. For both the $N_f=8$ and $N_f=2$ sextet theories, we find the dominant contributions to be
\begin{align}
\frac{\Delta M^2_\pi}{M^2_\pi} & \sim  \frac{M^2_\pi}{N_f(4\pi F_\pi)^2},\label{Eq:Loop1}\\
\frac{\Delta F_\pi}{F_\pi} & \sim \frac{M^2_\pi N_f}{2(4\pi F_\pi)^2},\label{Eq:Loop2}\\
\frac{\Delta M^2_d}{M^2_d} & \sim \frac{M^2_\pi (N^2_f-1)}{(4\pi F_d)^2}.\label{Eq:Loop3}
\end{align}
Each expression arises only from loops of NGB's. Expressions (\ref{Eq:Loop2}) and (\ref{Eq:Loop3}) can be large in the case of the $N_f = 8$ theory, clearly dominating contributions arising from virtual scalars proportional to $M_d^2/(4 \pi F_d)^2$. Expression (\ref{Eq:Loop2}) is familiar from chiral perturbation theory (in the absence of a dilaton).  Expression (\ref{Eq:Loop1}) is also familiar from chiral perturbation theory. It is small in the case of the $N_f = 8$ theory, roughly the same size as the scalar-loop contribution. It dominates this contribution in the case of the $N_f=2$ sextet theory, but remains small.

For the $N_f=8$ theory,
\begin{align}
\frac{\Delta M^2_\pi}{M^2_\pi} \lesssim 0.02,\qquad
\frac{\Delta F_\pi}{F_\pi} \lesssim 0.5,\qquad
\frac{\Delta M^2_d}{M^2_d} \lesssim 0.5,\;\;
\end{align}
where we have used Eq. (\ref{Eq:Ratio8}) to estimate the size of $F_d^2$. For the $N_f=2$ sextet theory,
\begin{align}
\frac{\Delta M^2_\pi}{M^2_\pi} \lesssim 0.05,\qquad
\frac{\Delta F_\pi}{F_\pi} \lesssim 0.1,\qquad
\frac{\Delta M^2_d}{M^2_d} \lesssim 0.03,
\end{align}
where we have used Eq.~(\ref{Eq:RatioS}) to estimate the size of $F^2_d$.

\subsection{Summary of Corrections}

Heavy-state corrections arise from scales $M_H$ on the order of the EFT cutoff $4 \pi F_{\pi}$. Our rough estimates indicate that they are no larger than $20 - 30 \%$. This is smaller than the errors quoted in Section~\ref{Sec:Lattice} for $f_{\pi}^2/f_d^2$. However, the heavy-state corrections  could be somewhat larger than the errors quoted there for $y$. The quantum loop corrections include NGB counting factors which are not large in the case of the $N_f=2$ sextet theory. We estimate the loop corrections for this theory to be no larger than $ 10 \%$.  For the $N_f = 8$ theory, however, they could be larger in the case of $\Delta F_{\pi}/F_{\pi}$ and $\Delta M_{d}^2/M_{d}^2$. We can bound them only at approximately the $50\%$ level.
 
Despite these uncertainties, the classical EFT has provided a successful fit with a small $\chi^2/N_{dof}$ to the smoothly varying lattice data for both theories. This suggests that the  full set of corrections is relatively small even for the $N_f = 8$ theory.  Also, it is important to note that since the corrections depend only on \textit{ratios} of capitalized quantities, they show relatively little variation as a function of $m$. That is, their systematic effect is expected to be even smaller. A key question is whether the 5\% determination of the $y$ parameter based on Eqs.~(\ref{Eq:LSDy}) and (\ref{Eq:Kutiy}) is reliable, especially for the $N_f = 8$ theory. The quality of the statistical fits we performed with the classical EFT suggests that it \textit{is} reliable and that the error estimates of this Section should be taken to be conservative, but this issue requires further analysis.

\section{Conclusions}
\label{Sec:conclusions}

We have developed and analyzed a dilaton-based EFT for the description of lattice data for gauge theories in which the number of massless fermions is near but below the transition from confinement to infrared conformal behavior. We have applied it here to $SU(3)$ gauge theories with $N_f = 8$ Dirac fermions in 
the fundamental representation~\cite{LSD,LatKMI}, and with $N_f = 2$ Dirac fermions in the $2$-index symmetric (sextet) representation~\cite{FHKNSW,FHKMNW,FHKMNW2}. Both show evidence for the existence of a remarkably light scalar particle. 

The EFT incorporates the scalar particle and the approximate NGB's associated with the spontaneous breaking of chiral symmetry. 
It includes a dilaton potential describing small explicit breaking of conformal symmetry. It also includes an NGB mass, 
explicitly breaking the chiral symmetry, arising from the presence of the underlying fermion mass $m$, necessary for lattice simulations. 
We have shown that the EFT applied at the classical level accurately describes the existing lattice data as a function
of the fermion mass, and that the EFT can be conveniently expressed in terms of measured, finite-$m$ quantities (denoted by capital letters) as in Eqs.~(\ref{Eq:LpiR}) - (\ref{Eq:Wexpansion}).
The EFT naturally accommodates the fact that $F_{\pi}^2$ varies substantially and linearly as a function of $m$
throughout its range. Our fits of the classical theory to the data have determined the parameter $y$, taken to be a constant, at the $5\%$ level. This 
parameter has been identified with the scaling dimension of $\bar{\psi}\psi$ \cite{LLB}. For both theories discussed here, we find a value consistent with $y = 2$.

The substantial variation of the measured quantities with $m$ in the lattice data suggests that these quantities extrapolate to even smaller values in the $m \rightarrow 0$ limit. Thus the mass term in Eq.~(\ref{Eq:LM}) should be regarded as a significant deformation of the $m = 0$ EFT in the range of the data. The $y$ parameter is a property of this deformation and can therefore be well constrained. Due to the deformation, we can probe the potential $V(\chi)$ at field values well above its minimum. We have concluded that for both theories, $V(\chi)$ grows approximately as $\chi^4$ in this range (the deviation from $4$ in the exponent $p$ is very similar for the two theories). We have also provided a determination of the ratio $f_{\pi}^2/f_d^2$ since within our framework, it is related to quantities defined away from the $m \rightarrow 0$ limit. Determining other extrapolated quantities, for example the ratio $m_d^2/(4 \pi f_{d})^2$, will require data at smaller values of $m$ and a knowledge of the potential $V(\chi)$ in the neighborhood of its minimum. Nevertheless, the trend of the lattice data as $m$ decreases suggests that this ratio is small, and that in this limit the scalar \textit{is} a dilaton. 

For both theories, we have found  $f_d/f_\pi\sim3$, leading to the prediction $F_d \sim 3 F_{\pi}$ throughout the range of $m$ values. This is a testable result, as the decay constant of the dilaton can be measured in future lattice calculations. The size of the ratio $f_d/f_\pi$ suggests the presence of additional condensates besides the chiral condensate, adding support to the interpretation of the light scalar particle as a dilaton.  The fact that $f_d/f_\pi\sim3$ also implies that the two models discussed in this paper require further extension if they are to replace the standard-model interpretation of the scalar particle of mass 125 GeV discovered by the LHC collaborations~\cite{Aad:2012tfa,Chatrchyan:2012xdj}. Measurements of the $pp \rightarrow h \rightarrow WW$ rate bound the vacuum value of the scalar to be close to the electroweak-symmetry-breaking scale~\cite{TheATLASandCMSCollaborations:2015bln}.

The classical EFT remains weakly coupled throughout the range of the data, meaning that the capitalized quantities satisfy $M_{\pi}^{2}, M_{d}^2  \ll (4 \pi F_{\pi})^2$. As a consequence, the EFT interprets the pseudoscalars as NGB's throughout the data range. This interpretation could break down at still larger values of $m$ where the breaking of the underlying chiral symmetry becomes dominantly explicit (as in conformal perturbation theory \cite{CPT}). We have estimated corrections to the classical EFT arising from heavy particles and from quantum loop corrections computed within the EFT. We have found that the heavy particle corrections are no larger than $20-30\%$. Similar remarks apply to our estimates of quantum loop corrections, although NGB counting factors led there to a weaker upper bound (as much as $50\%$) in the case of the $N_f=8$ theory. Since the effective cutoff $M_H\sim4 \pi F_{\pi}$ scales with $m$ in the same way as $M_\pi$ and $M_d$, the systematic effect should be smaller.

Our investigation motivates further lattice studies of confining gauge theories near the edge of the conformal window. For the $N_f=8$ and sextet $N_f=2$ theories, taking data at smaller values of $m$ will eventually reveal the functional form of the scalar potential $V(\chi)$ at field values near its minimum, enabling a determination of parameters $m^2_d,\,f^2_d$ and $f^2_\pi$. Taking data in the current $m$ range for additional observables, such as $\pi\pi$-scattering lengths, will test predictions made by the EFT and therefore check the consistency of the framework. We find it fascinating that the simple EFT studied here accurately describes two different theories relatively close to the conformal window, and does so with parameters $y$ and $p$ so close in magnitude. Lattice studies of other similar gauge theories will be important to further test the generality of the dilaton EFT.


\vspace{1.0cm}
\begin{acknowledgments}
We thank Julius Kuti, Robert Shrock, George Fleming, Andrew Gasbarro, David Schaich and the members of the LSD collaboration for helpful discussions. The work of MP has been supported in part by the STFC Consolidated Grants ST/L000369/1 and ST/P00055X/1.
\end{acknowledgments}

\vspace{1.0cm}
\appendix
\appendix

\end{document}